\documentclass[aps,prb,reprint,showpacs,superscriptaddress,longbibliography]{revtex4-2}
\usepackage{mathtools,amssymb,graphicx,units}
\usepackage[usenames,dvipsnames]{color}
\usepackage[plainpages=false,pdfpagelabels,colorlinks=true,linkcolor=red,urlcolor=blue,citecolor=blue,pdftitle={Title},pdfauthor={},pdfdisplaydoctitle=true,pdfduplex=DuplexFlipLongEdge]{hyperref}
\usepackage{subfigure}
\usepackage{grffile}
\usepackage{bm}
\usepackage{mhchem}
\usepackage[normalem]{ulem}
\usepackage{hyperref}
\usepackage{bibentry}
\usepackage{natbib}

\newcommand{\abs}[1]{\ensuremath{\left| #1 \right|}}
\newcommand{\ket}[1]{\left\vert#1\right\rangle}

\newcommand{\braket}[2]{\left\langle#1\left\vert#2\right.\right\rangle}

\newcommand{\vev}[1]{\left\langle #1\right\rangle}
\newcommand{\1}{\mbox{\bf 1}}
\newcommand{\ud}{\mathrm{d}}

\newcommand\redsout{\bgroup\markoverwith{\textcolor{red}{\rule[0.5ex]{2pt}{0.4pt}}}\ULon}

\begin{document}
\title{Topological phase in the extended Haldane-Hubbard model with sublattice-dependent repulsion}

\author{Bao-Qing Wang}
% \affiliation{Lanzhou Center for Theoretical Physics, Key Laboratory of Theoretical Physics of Gansu Province, and Key Laboratory of Quantum Theory and Applications of MoE, Lanzhou University, Lanzhou, Gansu 730000, China}
\affiliation{Key Laboratory of Quantum Theory and Applications of MoE, Lanzhou Center for Theoretical Physics, and Key Laboratory of Theoretical Physics of Gansu Province, Lanzhou University, Lanzhou 730000, China}

\author{Can Shao}
\email{shaocan@njust.edu.cn}
\affiliation{Department of Applied Physics, Nanjing University of Science and Technology, Nanjing 210094, China}

\author{Takami Tohyama}
\affiliation{Department of Applied Physics, Tokyo University of Science, Tokyo 125-8585, Japan}

\author{Hong-Gang Luo}
\affiliation{Key Laboratory of Quantum Theory and Applications of MoE, Lanzhou Center for Theoretical Physics, and Key Laboratory of Theoretical Physics of Gansu Province, Lanzhou University, Lanzhou 730000, China}
\affiliation{Beijing Computational Science Research Center, Beijing 100084, China}

\author{Hantao Lu}
\email{luht@lzu.edu.cn}
\affiliation{Key Laboratory of Quantum Theory and Applications of MoE, Lanzhou Center for Theoretical Physics, and Key Laboratory of Theoretical Physics of Gansu Province, Lanzhou University, Lanzhou 730000, China}

\date{\today}

\begin{abstract}
We study the ground-state phase diagram of the half-filled extended Haldane-Hubbard model on the honeycomb lattice with sublattice-dependent on-site repulsion ($U_{\text{A/B}}$) using the exact diagonalization (ED) and mean-field (MF) methods. The resulting phase diagram shows that there is a topologically nontrivial phase with the Chern number $C=1$, emerging via the development of the imbalance between $U_{\text{A}}$ and $U_{\text{B}}$. In this phase, the antiferromagnetic correlations are observed in the ED calculation, in line with the finite antiferromagnetic order obtained by the MF method. The spontaneous symmetry breaking of SU(2) spin rotation in the phase is also identified in the MF level. Distinct from previous studies in which the exotic $C=1$ phase relies on the interplay between sublattice-dependent potentials and electronic interactions, our paper presents an alternative way by solely tuning the on-site interactions.
\end{abstract}

\maketitle

% \paragraph{Introduction.---}
\section{Introduction}
\label{sec_intr}

Presently quantum phase transitions (QPTs) of condensed matter can be categorized into two paradigms. The first paradigm, described by the Ginzburg-Landau theory, involves spontaneous symmetry breaking and is characterized by local order parameters~\cite{Landau_1980}. The second paradigm comprises topological phase transitions, which, identified by proper topological invariants, have been classified quite completely in noninteracting systems under various symmetries~\cite{Kruthoff_2017,Zhang_19,Vergniory_19,Tang_19}. Naturally exotic intermediate states can be expected in the presence of competition between local and topological orders. However in the case of correlated systems, local orders are usually driven by interactions, and the nature of the intermediate states could be elusive.

In the study of the interplay between topology and interactions in condensed matter systems~\cite{Rachel_2018,Hohenadler_2013}, a promising direction is to consider systems with competitive local orders under interactions, while in the absence of interactions they are in topologically nontrivial phases. A prominent example is the Haldane model on honeycomb lattice~\cite{Haldane_1988} plus various interactions. For instance, studies on the phase diagrams of the spinless Haldane model with nearest-neighbor (NN) interactions reveal a coexistence of topological and charge-density-wave (CDW) orders, albeit due to finite-size effects~\cite{Varney_2010,Varney_2011}. In the spinful Haldane-Hubbard model, an intriguing phase (with Chern number $C=1$) emerges where one spin species resides in a topologically non-trivial state while the other remains localized~\cite{He_2011}. Results from various methods, including mean field (MF), exact diagonalization (ED), quantum Monte Carlo, density matrix renormalization group (DMRG), and dynamical mean-field theory (DMFT), suggest that the spontaneous spin symmetry breaking arises mainly from the competition between sublattice potentials and on-site Coulomb repulsions~\cite{He_2011,Zhu_2014,Vanhala_2016,Tupitsyn_2019,Mertz_2019,Shao_2023,He_2023}. Moreover, sublattice potentials appear essential to this exotic state, with other factors exhibiting flexibility in one way or another. This flexibility is evident in phenomena with $C=1$ observed on square lattices instead of honeycomb lattices~\cite{Wang_2019, Ebrahimkhas_2021}, in models like the Kane-Mele-Hubbard instead of the Haldane-Hubbard model~\cite{Jiang_2018}, in the presence of double exchange processes rather than Coulomb interactions~\cite{Tran_2022}, and in the introduction of disordered potentials to the interacting Haldane system~\cite{Silva_2024}.

In this paper, we proposed that the exotic $C=1$ state can be accessed solely through the interplay between topology and interactions, without relying on the assistance of sublattice potentials. We consider the Haldane-Hubbard model with sublattice-dependent Coulomb repulsions, i.e., $U_{\text{A}}$ and $U_{\text{B}}$, and study its ground-state phase diagram. For both MF and ED methods, the $C=1$ state is observed, although its occurrence is narrower in the ED results. However, incorporating the NN interaction $V$ into the model can stabilize the $C=1$ phase observed in the ED calculations. Other phases include the CDW phase governed by $V$, the spin-density-wave state (SDW) governed by $U$, and the Chern insulator (CI) phase with $C=2$ under weak interactions. Except for the Chern number, phase transitions are also identified by structure factors of the SDW and CDW, and the results suggest a topological antiferromagnetic state for the $C=1$ phase. By analyzing the effective Dirac masses in the MF results, its origin can be also attributed to the spontaneously broken spin rotation symmetry. 
% \tred{We suggest that the experimental realization of this model may be easier compared to those with additional potentials, especially in cold atom systems.}

% \paragraph{Model and observables.---}
\section{Model and observables}
\label{sec_model}
% site-dependent Coulomb repulsion staggered ? 
% spatial modulation of the interaction 
% spatially alternating interactions
% in close comparison 
% antiferromagnetic Chern insulator
In this article, an extended Haldane-Hubbard model for electrons on the honeycomb lattice with sublattice-dependent Coulomb repulsion is studied. The Hamiltonian reads
\begin{equation}
\hat{H}=\hat{H}_{0}+\hat{H}_{I},
\label{eq_ham}
\end{equation}
with the kinetic part and the interaction part as
\begin{eqnarray}
\hat{H}_{0}=&-&t_{1}\sum\limits_{\vev{ij},\sigma}\left({c}_{i\sigma}^{\dag}{c}_{j\sigma}+\text{H.c.}\right)\nonumber \\
&-&t_{2}\sum\limits_{\vev{\vev{ij}},\sigma}\left(e^{\mathrm{i}\phi_{ij}}{c}_{i\sigma}^{\dag}{c}_{j\sigma}+\text{H.c.}\right),
\label{eq_ham1}
\end{eqnarray}
and
\begin{eqnarray}
\hat{H}_{\text{I}}=
% \sum\limits_{i,\alpha\in\{\text{A},\text{B}\}}
\sum\limits_{i,\alpha}
U_{\alpha}\hat{n}_{i\uparrow}\hat{n}_{i\downarrow}+V\sum\limits_{\vev{ij},\sigma\sigma^{\prime}}\hat{n}_{i\sigma}\hat{n}_{j\sigma^{\prime}}.
\label{eq_ham2}
\end{eqnarray}
Here ${c}_{i\sigma}^{\dagger}$ (${c}_{i\sigma}$) represents the creation (annihilation) operator of an electron with spin $\sigma$ ($\sigma=\uparrow,\,\downarrow$) at site $i$, and the corresponding number operator $\hat{n}_{i\sigma}={c}^{\dagger}_{i\sigma}{c}_{i\sigma}$. The NN and the next-nearest-neighbor (NNN) hopping amplitudes are $t_1$ and $t_2$, respectively. In the NNN hopping terms, the Haldane phase $\phi_{ij}$ is added with the absolute value fixed and a sign depending on the clockwise (anticlockwise) direction of the loops inside the honeycombs~\cite{Haldane_1988}. The phase breaks the time reversal symmetry explicitly without net magnetic flux. Among the interaction terms, $U_{\alpha}$ denotes the on-site Coulomb repulsion strength in the sublattice $\alpha$ ($=\text{A},\text{B}$), and $V$ is the NN interaction in the extended Hubbard model (e.g., see Refs.~\cite{Baeriswyl_1985,Lin_2000}). Note that instead of the staggered potentials in the previous studies~\cite{He_2011,Zhu_2014,Vanhala_2016,Tupitsyn_2019,Mertz_2019,Shao_2023,He_2023}, the on-site interaction $U_{\alpha}$ here is considered to be sublattice dependent, which also breaks the sublattice (inversion) symmetry explicitly.

\begin{figure}
\centering
\includegraphics[width=0.45\textwidth]{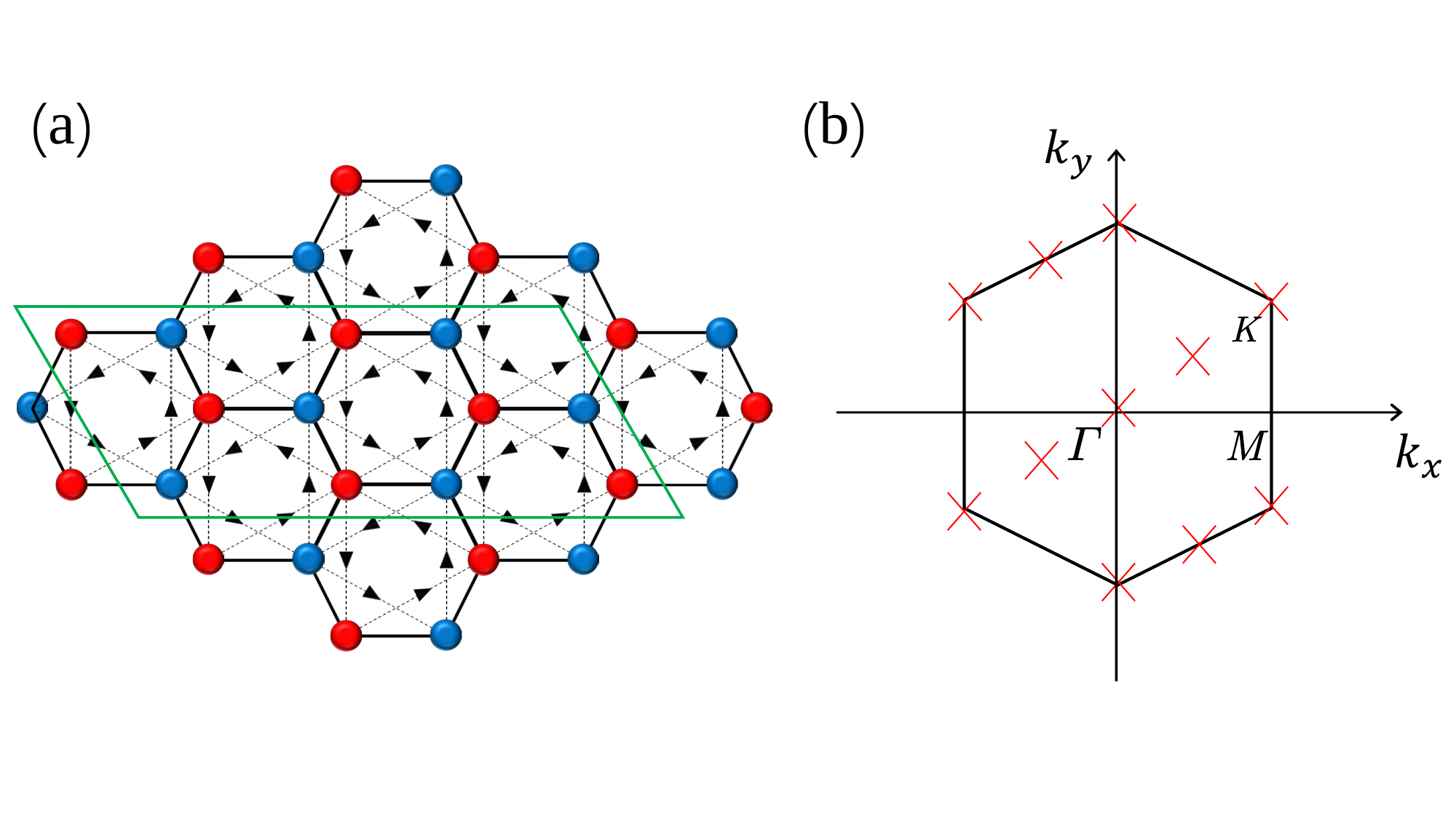}
\caption{(a) A schematic plot of a honeycomb lattice with two triangular sublattices distinguished by different colors. The arrows indicate the directions of the positive phase winding for the complex NNN hoppings. The 12A cluster used in the ED calculation is indicated by the parallelogram. (b) The set of $k$ points (cross symbols) in the momentum space for the 12A cluster.}
\label{fig_lattice}
\end{figure}

In the present study, we focus on the effect of the sublattice-dependent on-site interaction on the phase diagram of the extended Haldane-Hubbard model at half-filling. Special attention is paid to the topologically nontrivial phases. The phase diagram as a function of the on-site interaction $U_{\alpha}\ (\alpha=\text{A},\text{B})$ and the NN interaction $V$ is investigated by the ED method, complemented by the MF calculations. To mitigate the finite-size effect and to avoid overlooking vital information in the low-lying spectra, it is important to choose proper clusters in the ED calculation on two-dimensional lattices. In our case, whether the reciprocal lattice includes the $K$ points can be essential~\cite{Varney_2010,Varney_2011,Shao_2023}. Here we chose a cluster including 12 lattice sites [Fig.~\ref{fig_lattice}(a)] under the periodic boundary condition (PBC). Compared to other clusters with size up to $18$, the reciprocal lattice of the so-called 12A cluster [Fig.~\ref{fig_lattice}(b)] incorporates more high-symmetry points, including the $\Gamma$ point, all the $K$ points and one pair of $M$ points~\cite{Shao_2021}. For a cluster with better symmetries, we have to go to as far as $24$ sites, which exceeds our current computational resources. 

In the ED calculation, to specify different phases in the phase diagram, both the local orders, including the SDW and CDW, and the topological invariant, i.e., Chern number here, are evaluated. The SDW and CDW orders are the ones that dominate in the large-$U$ and large-$V$ limits, respectively. Their structure factors are written in a staggered fashion as
\begin{eqnarray}
S_{\text{SDW}}=\frac{1}{N}\sum_{i,j}(-1)^{\eta}\vev{\left(\hat{n}_{i\uparrow}-\hat{n}_{i\downarrow}\right)\left(\hat{n}_{j\uparrow}-\hat{n}_{j\downarrow}\right)},\nonumber\\
S_{\text{CDW}}=\frac{1}{N}\sum_{i,j}(-1)^{\eta}\vev{\left(\hat{n}_{i\uparrow}+\hat{n}_{i\downarrow}\right)\left(\hat{n}_{j\uparrow}+\hat{n}_{j\downarrow}\right)},
\label{eq_SDWCDW}
\end{eqnarray}
where $\eta=0\ (\eta=1)$ if sites $i$ and $j$ are in the same (different) sublattice. 

The Chern number of a many-body state can be calculated by the integration of the Berry curvature of the corresponding wavefunction over the twisted boundaries~\cite{Niu_1985}
\begin{equation}
C=\iint\frac{\ud\theta_{x}\,\ud\theta_{y}}{2\pi i}
\left[\braket{\frac{\partial\Psi}{\partial{\theta_{x}}}}{\frac{\partial\Psi}{\partial\theta_{y}}}
-\braket{\frac{\partial\Psi}{\partial\theta_{y}}}{\frac{\partial\Psi}{\partial\theta_{x}}}\right],
\label{eq_Chern}
\end{equation}
where $\ket{\Psi}$ is the many-body wavefunction, and $\theta_{x}(\theta_y)\in[0,2\pi)$ is the twisted angle along the $x\ (y)$ direction in the boundary condition. In our calculation, a mesh of ${12\times 12}$ in the $(\theta_x,\theta_y)$ space is sufficient to guarantee the convergence~\cite{Kudo_2019,Varney_2011,Shao_2023}.

On the other hand in the MF calculation, the SDW and CDW order parameters are defined as
\begin{eqnarray}
\mathcal{O}_{\text{SDW}}&=&\abs{\frac{1}{2}\left(\vev{\vec{S}_{\text{A}}}-\vev{\vec{S}_{\text{B}}}\right)},\nonumber\\
\mathcal{O}_{\text{CDW}}&=&\abs{\left(n^{\text{A}}_{\uparrow}+n^{\text{A}}_{\downarrow}\right)-\left(n^{\text{B}}_{\uparrow}+n^{\text{B}}_{\downarrow}\right)},
\end{eqnarray}
where $\vec{S}_{\text{A}(\text{B})}$ is the MF magnetization vector for the A(B) sublattice, $n^{\text{A}}_{\uparrow}$ is the MF density of spin-up electrons on A-sublattice, etc. More details of the MF method and results can be found in Appendix~\ref{app_MF_in_moment}.
% $\vec{S}_{i}=\sum_{\alpha\beta}c_{i\alpha}^{\dagger}\vec{\sigma}_{\alpha\beta}c_{i\beta}$ ($\vec{\sigma}$ is the vector of three Pauli matrices)

In the following, we set the reduced Planck constant $\hbar=1$, and $t_1$ as the energy unit. The NNN hopping amplitude $t_2$ and the phase amplitude $\phi$ are fixed to be $0.2$ and $\pi/2$, respectively. That is, without interactions, the ground state of the system is in the CI phase with $C=2$. 

% \paragraph{Results.---}
\section{Results}
\label{sec_results}

\begin{figure}
\centering
\includegraphics[width=0.5\textwidth]{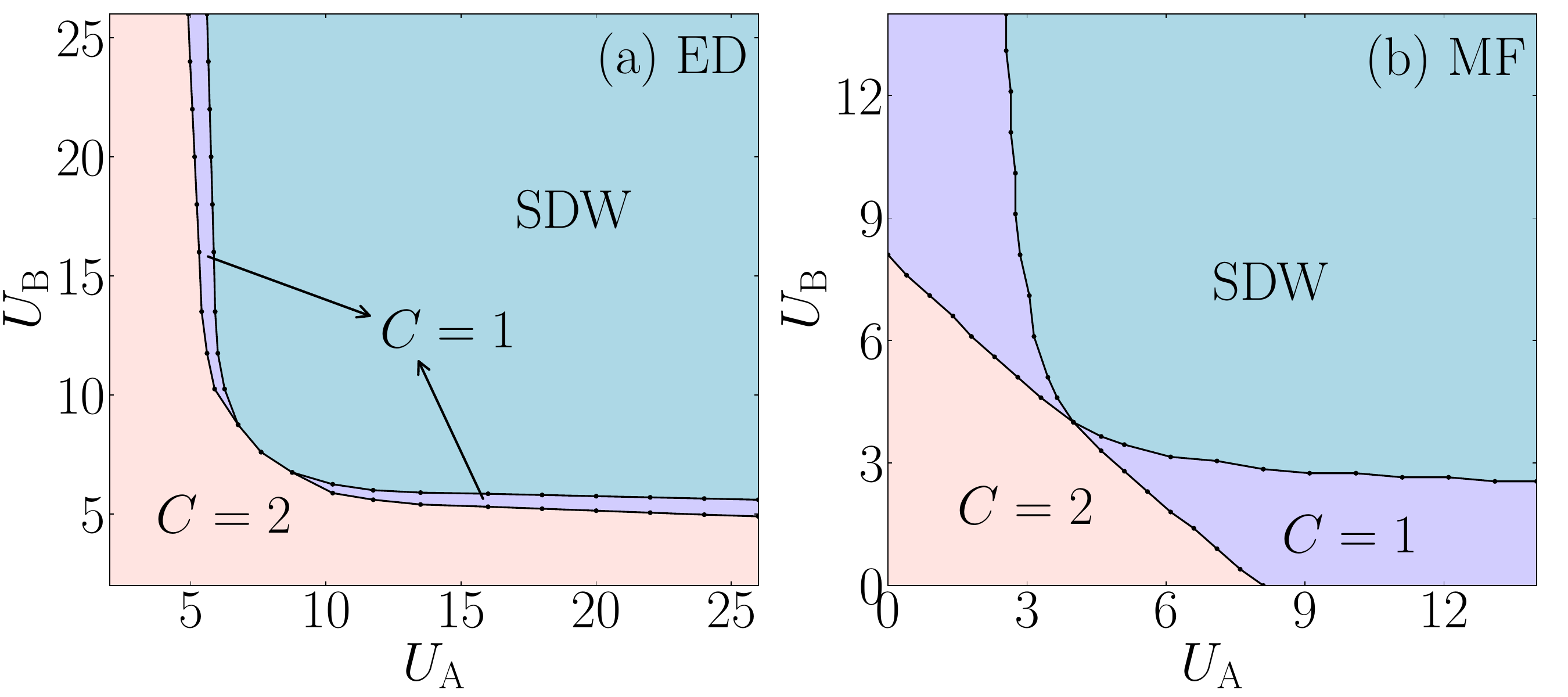}
\caption{The phase diagram of the Hamiltonian~(\ref{eq_ham}) with $V=0$ as a function of alternating on-site interactions $U_{\text{A}}$ and $U_{\text{B}}$, obtained by ED (a) and MF (b), respectively. In the absence of the NN interaction $V$, three phases are identified: the $C=2$ Chern-insulator phase, the topologically trivial SDW phase, and sandwiched between them, the $C=1$ AFCI phase.}
\label{fig_v0}
\end{figure}

The essential result of the study is presented in Fig.~\ref{fig_v0}, where the phase diagrams of the Hamiltonian~(\ref{eq_ham}) obtained by the ED [on the 12A cluster (see Fig.~\ref{fig_lattice})] and MF in the ($U_{\text{A}}$, $U_{\text{B}}$) space without the NN interaction are shown. The phase boundaries are determined by the structure factors (or the order parameters in MF) and the Chern numbers together. We see that with the sublattice symmetry being {\em explicitly} broken by the spatial alternation of the on-site interaction, a $C=1$ phase emerges. The phase is sandwiched between the $C=2$ Chern insulator and the topologically trivial SDW phase and expands with the increase of the imbalance between $U_{\text{A}}$ and $U_{\text{B}}$. We note that the MF results have already converged on the $30\times 30$ lattice, and both the ED and MF calculations support the existence of the $C=1$ phase. Nevertheless, in the ED-produced phase diagram [Fig.~\ref{fig_v0}(a)], the region occupied by the phase is quite limited, and cannot extend into the small $U_{\text{A/B}}$ area, which is different from the MF result [Fig.~\ref{fig_v0}(b)]. Although the finite-size effects are expected to be moderate here due to the nonvanishing of the band gap of the noninteracting Hamiltonian (which is different from the case with staggered potential, e.g., in Ref.~[\onlinecite{Vanhala_2016}]), we suggest that in the future study further investigations are required for the existence and the exact location of the $C=1$ phase in the $U_\text{A}$-$U_\text{B}$ phase diagram (with $V=0$).

\begin{figure}
\centering
\includegraphics[width=0.5\textwidth]{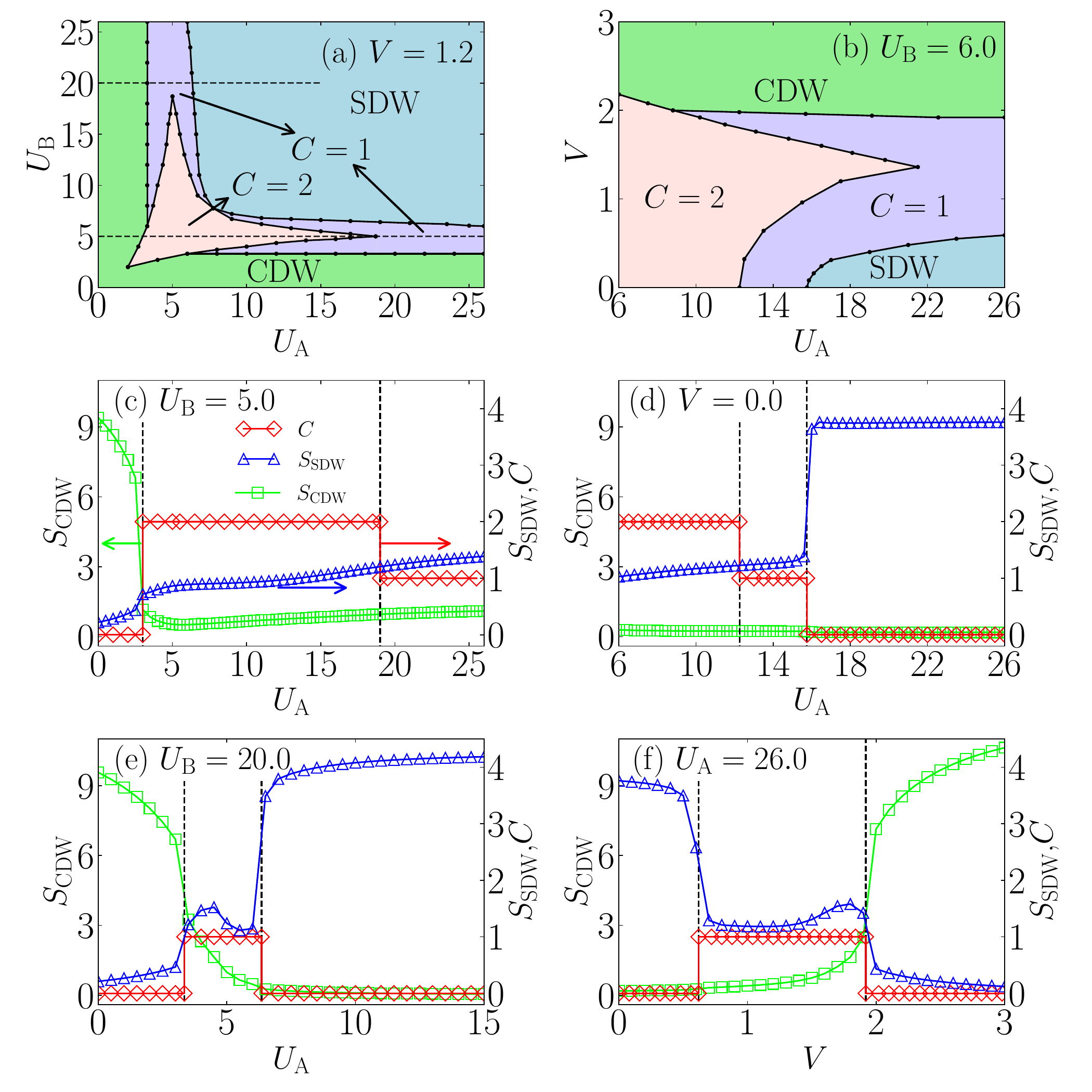}
\caption{(Top row) The phase diagrams of the model~(\ref{eq_ham}) obtained by ED for a fixed $V=1.2$ (a) and $U_{\text{B}}=6.0$ (b), respectively. (Bottom rows) The following two rows present the results of the CDW and SDW structure factors, i.e., $S_{\text{CDW}}$ (square symbols) and $S_{\text{SDW}}$ (triangular symbols), and the Chern number $C$ (diamond symbols) along the cuts (edges) in the phase diagrams. Among them, (c) and (e) for the two horizontal cuts in (a) at $U_{\text{B}}=5.0$ and $20.0$, respectively; (d) and (f) for the two edges in (b), one for $V=0.0$ (the lower horizontal edge) and the other for $U_{\text{A}}=26.0$ (the right-vertical edge). The vertical-dashed lines in (c)-(f) indicate the positions of the phase transition points on the cuts (edges).} 
\label{fig_ed}
\end{figure}

Up to now, we have provided evidence that apart from the ionic Haldane-Hubbard model where the sublattice (inverse) symmetry is broken at the single-particle level by the staggered potential, the Hamiltonian~(\ref{eq_ham}) that breaks the symmetry from the side of interaction also has the potential to support the $C=1$ phase. As will be addressed later in more detail, we furthermore observe the clear signatures of the developing SDW correlations in the phase (e.g., see Fig.~\ref{fig_ed}). As a result, following Refs.~\onlinecite{Jiang_2018,Wang_2019} we also call the $C=1$ phase the antiferromagnetic Chern insulator (AFCI) phase.

The nature of the AFCI phase here is indeed identical to that of the topological SDW state, which, to the best of our knowledge, was first suggested in Ref.~[\onlinecite{He_2011}] for the ionic Haldane-Hubbard model and dubbed ``B-TSDW". A physical description of the state is that due to a spontaneous spin-rotation symmetry breaking caused by the interplay of topology and the electronic interactions, one of the spin components remains topologically nontrivial whereas the other does not. In our case, we can easily understand on the MF level that the inequality of the on-site interaction concerning (A/B) sublattice can apparently produce an effective staggered potential. Together with the magnetic effect of the electron-electron interactions in general, a {\em spin-dependent} staggered potential can appear, which, consequently, makes the emergence of the $C=1$ phase possible. More discussions of the issue on the MF level, including the evaluation of the Chern number in terms of the effective Dirac masses, can be found in Appendix~\ref{app_effective_mass}.

% transition points: (c) UA=3, 19; (e) UA=3.4, 6.4; (d) UA=12.25, 15.75; (f) V=0.62, 1.91 

In the following we show that the presence of the NN interaction $V$ can significantly modify the phase diagram in Fig.~\ref{fig_v0} and can help to stabilize the AFCI phase for wider range of $U_{\text{A/B}}$. There have been discussions on the possibility of inducing the AFCI phase via the NN interaction $V$, without explicitly breaking the sublattice symmetry~\cite{Shao_2021}. However, the existence of the phase there was found to be elusive. It remains to be seen whether the $V$ term that favors the charge order can play a similar role of, say, the staggered potential and induce the exotic AFCI phase. In our present model to address the effect of the NN interaction with the alternating on-site repulsion, we present the phase diagrams for several nonzero $V$s obtained by the ED method: one is in the $(U_{\text{A}},U_{\text{B}})$ space with fixed $V=1.2$ [Fig.~\ref{fig_ed}(a)], the other in the $(V,U_{\text{A}})$ space with fixed $U_{\text{B}}=6.0$ [Fig.~\ref{fig_ed}(b)]. The following two rows in Fig.~\ref{fig_ed} show the results of the structure factors defined by Eq.~(\ref{eq_SDWCDW}) and the Chern numbers along the cuts (or edges) in the corresponding phase diagrams.

In the phase diagrams, the presence of $V$ produces a new phase that is previously absent, i.e., the phase with prominent CDW correlations. Besides, the competition between the $V$-favored CDW order and the $U$-favored SDW order, both topologically trivial, can provide ample space to accommodate the topologically nontrivial phases, including $C=2$ and $C=1$ phases. From Fig.~\ref{fig_ed}(a) with $V=1.2$, we see that compared to the $V=0$ case [Fig.~\ref{fig_v0}(a)], the AFCI region is significantly enlarged in the presence of the NN interaction. The enhancement effect can also be identified in the phase diagram in the space of $(V,U_{\text{A}})$ with $U_{\text{A}}\ge U_{\text{B}}=3.0$ [Fig.~\ref{fig_ed}(b)], where we can observe that the turning on of $V$ leads to the suppression of the SDW correlations, and as a consequence, extend the reach of the topological phases. 

\begin{figure}
\centering
\includegraphics[width=0.5\textwidth]{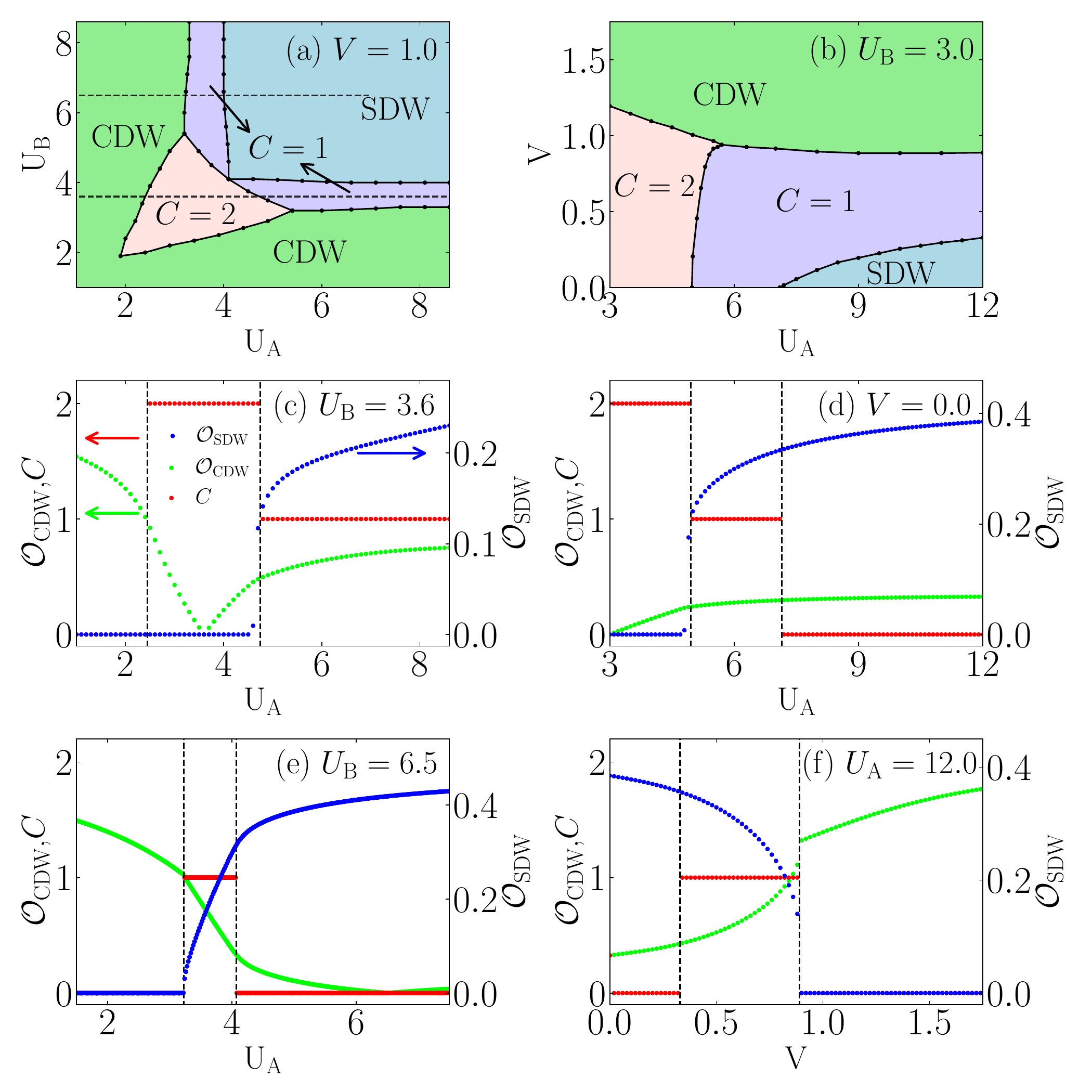}
\caption{(Top row) The phase diagrams of the model~(\ref{eq_ham}) obtained by the MF method on $360\times 360$ lattice for a fixed $V=1.0$ (a) and $U_{\text{B}}=3.0$ (b), respectively. (Bottom rows) The following two rows present the results of the CDW and SDW order parameters, i.e., $\mathcal{O}_{\text{CDW}}$ (in green) and $\mathcal{O}_{\text{SDW}}$ (in blue), and the Chern number $C$ (in red) along the cuts (edges) in the phase diagrams. Among them, (c) and (e) for the two horizontal cuts in (a) at $U_{\text{B}}=3.6$ and $6.5$, respectively; (d) and (f) for the two edges in (b), one for $V=0.0$ (the lower horizontal edge) and the other for $U_{\text{A}}=12$ (the right-vertical edge). The vertical-dashed lines in (c)-(f) indicate the positions of the phase transition points on the cuts (edges). Note that in (a) to reduce the area occupied by the CDW phase, the minimal values of $U_{\text{A}}$ and $U_{\text{B}}$ are both set to be $1.0$.}
\label{fig_mf}
\end{figure}

The nature of the phase transitions between the topologically trivial and nontrivial phases can be readily accessed from the evolution of the structure factors. For example, in Figs.~\ref{fig_ed}(e) and (f) where the cuts involve phase transitions between the AFCI with topologically trivial CDW and SDW phases, characteristic discontinuities in the structure factors at the transition points are observed. The observation is consistent with the previous studies~\cite{Varney_2010,Imriska_2016,Shao_2021,Shao_2023}, and we suggest that these quantum phase transitions are first-order ones. On the other hand, for the phase transitions between the topological phases, i.e., the AFCI and CI phases here, the evolution of the structure factors is quite smooth [e.g., see Figs.~\ref{fig_ed}(c) and (d)]~\cite{Shi_2021,Shao_2023}. The competition between the local orders also manifests itself in the topological phases as the remnant CDW and SDW correlations [Figs.~\ref{fig_ed}(c)-\ref{fig_ed}(f)]. It would be desirable to note the clear development of the SDW correlations in the AFCI phase, which indicates the coexistence of the local magnetic order and the global topological structures. The phenomenon can also be observed in the MF results [see Figs.~\ref{fig_mf}(c)-\ref{fig_mf}(f)].

% along the diagonal line, U=1.9; U=4.1
% transition points: triple point: UA=5.7, V=0.94;
% (c) UA=2.45, 4.75; (e) UA=3.23, 4.07; (d) UA=4.95, 7.15; (f) V=0.33, 0.89 

The results of the complementary Hartree-Fock MF study are presented in Fig.~\ref{fig_mf}. Similar to Fig.~\ref{fig_ed}, the first row shows phase diagrams in the $(U_{\text{A}},U_{\text{B}})$ space with fixed $V=1.0$, and in the $(V,U_{\text{A}})$ space with fixed $U_{\text{B}}=3.0$. The next two rows present the MF results of the local order parameters and Chern numbers along the cuts (or edges) in the corresponding phase diagrams. The Chern numbers are calculated in discretized Brillouin zones following the method in Ref.~[\onlinecite{Fukui_2005}]. 

Comparing the phase diagrams obtained by MF (Fig.~\ref{fig_mf}) to those obtained by ED (Fig.~\ref{fig_ed}), we see that besides quantitative deviations between the phase boundaries and critical values, there is little qualitative difference between them. This consistency indicates the reliability of the phase diagram as a whole, particularly concerning the existence of the AFCI phase. Nevertheless, one point we would like to underline is that in the AFCI phase, both the CDW and SDW orders remain similar to the ED analysis; while in the CI phase, unlike the ED result, the SDW order parameter $\mathcal{O}_{\text{SDW}}$ is suppressed [e.g., see Figs.~\ref{fig_mf}(c) and \ref{fig_mf}(d)].

% \paragraph{Summary and discussion.---}
\section{Summary and discussion}
\label{sec_summary}

In summary, we investigated the ground-state phase diagram of the spinful Haldane-Hubbard model on the honeycomb lattice with sublattice-dependent Coulomb repulsions. The topological AFCI phase ($C=1$) with finite antiferromagnetic order was identified via the ED and MF methods. The phase, achieved by solely tuning $U_{\text{A}}$ and $U_{\text{B}}$, can be further stabilized and expanded by the nearest-neighbor interaction $V$. This is distinguished from previous studies that rely basically on the interplay of sublattice-dependent potentials and the on-site interactions. The nature of the phase and the mechanism of its emergence can be understood qualitatively on the MF level.

In experiments, the Haldane model has been realized in cold atoms ten years ago~\cite{Jotzu_2014}, and more recently, in a solid-state material with moir{\`e} bilayers~\cite{Zhao_2024}. The realization of Hubbard interactions in fermionic cold atom systems has been reported~\cite{Esslinger_2010,Hart_2015,Messer_2015,Mazurenko_2017,Brown_2019,Nichols_2019}, including spatial modulation of the interaction in $\ce{^{174}Yb}$ gas systems~\cite{Yamazaki_2010}. This provides real prospects to emulate the extended Haldane-Hubbard model with site-dependent interactions in the near future. The physics of strongly correlated systems with site-alternating interactions can be rich and interesting~\cite{Li_2018,Gu_2019,Cheng_2024}, and we hope that the present paper can encourage more future studies.

% This suggests the potential fabrication of electron devices with dissipationless edge channels for specific spin directions in the future.

We note that recently there have been other proposals to realize the $C=1$ phase in the Haldane-Hubbard model by incorporating either the Anderson disorder~\cite{Silva_2024} or the spin-dependent sublattice potentials~\cite{Shao_2024}.

\begin{acknowledgments}
We thank Rubem Mondaini, Eduardo V. Castro, and Liang Du for the helpful discussions.
B.-Q.W and H.L. acknowledge support from the National Natural Science Foundation of China (NSFC; Grants No. 12174168, and No. 12247101). 
C.S. acknowledges support from NSFC (Grant No.12104229) and the Fundamental Research Funds for the Central Universities (Grant No. 30922010803).
H.-G.L. acknowledges funding from NSFC (Grants No. 11834005) and the National Key Research and Development Program of China (Grant No. 2022YFA1402704).
T.T. is partly supported by KAKENHI (Grant No. JP19H05825) from Ministry of Education, Culture, Sports, Science, and Technology, Japan.
\end{acknowledgments}

\appendix

\section{Mean-field method in momentum space}
\label{app_MF_in_moment}

The non-interacting part $\hat{H}_0$ of the Hamiltonian~(\ref{eq_ham}) in the main text can be written in the momentum space as 
\begin{eqnarray}
\hat{H_0}&=&\sum_{\mathbf{k},\sigma}
\left[\left(f^{\ast}(\mathbf{k})\,a_{\mathbf{k}\sigma}^{\dagger}b_{\mathbf{k}\sigma}+\text{H.c.}\right)\right.\nonumber\\
&&\left.+\left(m_{+}(\mathbf{k})a^{\dag}_{\mathbf{k}\sigma}a_{\mathbf{k}\sigma}+m_{-}(\mathbf{k})b^{\dag}_{\mathbf{k}\sigma}b_{\mathbf{k}\sigma}\right)
\right],
\label{eq_h0}
\end{eqnarray}
where the creation (annihilation) operator of electrons $c^{\dagger}$ ($c$) is denoted here as $a^{\dagger}$ ($a$) and $b^{\dagger}$ ($b$) for A and B sublattices, respectively. The coefficients $f$ and $m_{\pm}$ read
\begin{equation}
f(\mathbf{k})=t_1\left(1+2e^{ia\sqrt{3}k_y/2}\cos\frac{ak_x}{2}\right),
\label{eq_f} 
\end{equation}
\begin{eqnarray}
m_{+}(\mathbf{k})&=&-2t_2\left[\cos\left(\mathbf{k}\cdot\mathbf{a}_1-\phi\right)+\cos\left(\mathbf{k}\cdot\mathbf{a}_2+\phi\right)\right.\nonumber\\
&&\left.+\cos\left(\mathbf{k}\cdot\left(\mathbf{a}_1-\mathbf{a}_2\right)+\phi\right)\right],
\label{eq_m}
\end{eqnarray}
and $m_{-}(\mathbf{k})$ is obtained by simply replacing the Haldane phase $\phi$ in $m_{+}(\mathbf{k})$ with $-\phi$. The basis lattice vectors are set to be $\mathbf{a}_{1,2}={a}(\mp 1,\sqrt{3})/2$, where $a$ is the lattice spacing here.
% ~\footnote{The little abuse of the notation $a$ should arouse little confusion since the annihilation operator and the lattice constant appear in very different circumstances.}

The non-interacting Hamiltonian~(\ref{eq_h0}) can be put into a concise form as
\begin{equation}
\hat{H}_0=\sum_{\mathbf{k},\sigma}\left[a_{\mathbf{k}\sigma}^{\dagger}\ b_{\mathbf{k}\sigma}^{\dagger}\right]\mathcal{H}_0(\mathbf{k})
\begin{bmatrix}a_{\mathbf{k}\sigma}\\ b_{\mathbf{k}\sigma}\end{bmatrix}
\label{eq_h0b}
\end{equation}
with
\begin{equation}
\mathcal{H}_0(\mathbf{k})=C(\mathbf{k})\mathbf{1}+
\begin{pmatrix}
m(\mathbf{k}) & f^{\ast}(\mathbf{k}) \\ f(\mathbf{k}) & -m(\mathbf{k})
\end{pmatrix},
\label{eq_cmf}
\end{equation}
where $C(\mathbf{k})\pm m(\mathbf{k})=m_{\pm}(\mathbf{k})$. Note that at the valleys with coordinates as $\pm\mathbf{K}=\pm\frac{4\pi}{3a}(1,0)$, $f(\pm\mathbf{K})=0$, $C(\pm\mathbf{K})=3t_2\cos\phi$, and $m(\pm\mathbf{K})=\pm 3\sqrt{3}t_2\sin\phi$. We see that for the effective Dirac Hamiltonians in the vicinity of the valley points that are related by the spatial inversion, the masses $m(\pm\mathbf{K})$ have exactly {\em opposite} signs when $\phi\neq 0$. Consequently, the Chern number of the lower band (here temporally drop the spin degeneracy) can be either $1$ or $-1$, depending on the sign of $\sin\phi$. On the other hand, the system can be driven into the topologically trivial phase with zero Chern number by, e.g., introducing a staggered sublattice potential into the Hamiltonian as $\Delta\left(\hat{n}^{A}-\hat{n}^{B}\right)$, since with $\Delta>\abs{3\sqrt{3}t_2\sin\phi}$, the masses have equal signs.

Now we move to the interacting part $\hat{H}_\text{I}$ of the Hamiltonian (1) in the main text. It is approximated at the Hartree-Fock level in our mean-field (MF) treatment. Specifically, for the on-site interaction term (i.e. the $U$ term) we have
\begin{eqnarray}
\hat{n}_{i\uparrow}\hat{n}_{i\downarrow}&\simeq&\vev{\hat{n}_{i\uparrow}}\hat{n}_{i\downarrow}+\hat{n}_{i\uparrow}\vev{\hat{n}_{i\downarrow}}\nonumber \\
&-&\vev{{c}^{\dag}_{i\uparrow}{c}_{i\downarrow}}{c}^{\dag}_{i\downarrow}{c}_{i\uparrow}-\vev{{c}^{\dag}_{i\downarrow}{c}_{i\uparrow}}{c}^{\dag}_{i\uparrow}{c}_{i\downarrow}+\text{const};
\end{eqnarray}
For the NN interaction term (i.e. the $V$ term), 
\begin{eqnarray}
&&\hat{n}_{i\sigma}\hat{n}_{j\sigma^{\prime}}\simeq\vev{\hat{n}_{i\sigma}}\hat{n}_{j\sigma^{\prime}}+\hat{n}_{i\sigma}\vev{\hat{n}_{j\sigma^{\prime}}} \nonumber \\ 
&-&\vev{{c}^{\dag}_{i\sigma}{c}_{j\sigma^{\prime}}}{c}^{\dag}_{j\sigma^{\prime}}{c}_{i\sigma}
-{{c}^{\dag}_{i\sigma}{c}_{j\sigma^{\prime}}}\vev{{c}^{\dag}_{j\sigma^{\prime}}{c}_{i\sigma}}+\text{const}.
\end{eqnarray}
Depending on the lattice site, the creation (annihilation) operator $c^{\dagger}$ ($c$) in the above expressions can be either $a^{\dagger}$ ($a$) or $b^{\dagger}$ ($b$).

After the MF decoupling and the subsequent Fourier transformation, we finally obtain the MF Hamiltonian in the momentum space as following~\cite{Shao_2021}:
\begin{equation}
\hat{H}_{\text{MF}}=\hat{H}_{0}+\hat{H}_{\text{I-MF}},
\label{eq_hmf}	
\end{equation}
with 
\begin{equation}
\hat{H}_{\text{I-MF}}=\sum_{\mathbf{k}}\hat{\psi}_{\mathbf{k}}^{\dag}
\begin{pmatrix} \varepsilon_{\uparrow}^{\text{A}} & \xi_{\uparrow\uparrow}(\mathbf{k}) & \varepsilon_{\uparrow\downarrow}^{\text{A}} & \xi_{\uparrow\downarrow}(\mathbf{k}) \\ \xi_{\uparrow\uparrow}^{\ast}(\mathbf{k}) & \varepsilon_{\uparrow}^{\text{B}} & \xi_{\downarrow\uparrow}^{\ast}(\mathbf{k}) & \varepsilon_{\uparrow\downarrow}^{\text{B}} \\ \varepsilon_{\uparrow\downarrow}^{\ast \text{A}} & \xi_{\downarrow\uparrow}(\mathbf{k}) & \varepsilon_{\downarrow}^{\text{A}} & \xi_{\downarrow\downarrow}(\mathbf{k})   \\ \xi_{\uparrow\downarrow}^{\ast}(\mathbf{k}) & \varepsilon_{\uparrow\downarrow}^{\ast \text{B}} &  \xi_{\downarrow\downarrow}^{\ast}(\mathbf{k}) & \varepsilon_{\downarrow}^{\text{B}} \\
\end{pmatrix}\hat{\psi}_{\mathbf{k}},
\label{eq_himf}	
\end{equation}
where $\hat{\psi}^{\dag}_{\mathbf{k}}=\left[a^{\dag}_{\mathbf{k}\uparrow}\ b^{\dag}_{\mathbf{k}\uparrow}\ a^{\dag}_{\mathbf{k}\downarrow}\ b^{\dag}_{\mathbf{k}\downarrow}\right]$. The entries in the matrix read
\begin{align}
&\varepsilon_{\sigma}^{\text{A}}=U_{{\text{A}}}n^{\text{A}}_{-\sigma}+3V\sum_{\sigma^{\prime}}n^{\text{B}}_{\sigma^{\prime}},\nonumber\\
&\varepsilon_{\sigma}^{\text{B}}=U_{\text{B}}n^{\text{B}}_{-\sigma}+3V\sum_{\sigma^{\prime}}n^{\text{A}}_{\sigma^{\prime}},\nonumber\\
&\varepsilon_{\uparrow\downarrow}^{\text{A}}=-\frac{U_{\text{A}}}{N}\sum_{\mathbf{q}}\vev{a^{\dag}_{\mathbf{q}\downarrow}a_{\mathbf{q}\uparrow}}_{\text{MF}},\nonumber
\\ &\varepsilon_{\uparrow\downarrow}^{\text{B}}=-\frac{U_{\text{B}}}{N}\sum_{\mathbf{q}}\vev{b^{\dag}_{\mathbf{q}\downarrow}b_{\mathbf{q}\uparrow}}_{\text{MF}},\nonumber
\end{align}
\begin{align}
% &\varepsilon_{\uparrow\downarrow}^{\text{B}}=-\frac{U_{\text{B}}}{N}\sum_{\mathbf{q}}\vev{b^{\dag}_{\mathbf{q}\downarrow}b_{\mathbf{q}\uparrow}}_{\text{MF}},\nonumber\\
&\xi_{\sigma\sigma^{\prime}}(\mathbf{k})=-\frac{V}{N}\sum_{\mathbf{q}}g(\mathbf{k-q})\vev{b^{\dag}_{\mathbf{q}\sigma^{\prime}}a_{\mathbf{q}\sigma}}_{\text{MF}},
\label{eq_epsilon_and_xi}
\end{align}
where the function $g(\mathbf{k})=1+\exp({-i\mathbf{k}\cdot\mathbf{a}_1})+\exp(-i\mathbf{k}\cdot\mathbf{a}_2)$, and the density for the A-sublattice $n^{A}_{\sigma}=\frac{1}{N}\sum_{\mathbf{q}}\vev{a^{\dag}_{\mathbf{q}\sigma}a_{\mathbf{q}\sigma}}_{\text{MF}}$. Similar expression holds for $n^{B}_{\sigma}$. In our calculation, the averages $\vev{\cdots}$ are taken in the grand-canonical ensemble with respect to the MF Hamiltonian under sufficiently low temperatures. The chemical potential is also determined self-consistently by the filling. Numerically the convergence of the MF parameters can usually be guaranteed by the convergence of the free energy. To be brief, here we omit the expression of the free energy.

Once self-consistent MF solutions are found, the SDW and CDW order parameters can be computed. The Chern number for the occupied bands with respect to the MF Hamiltonian is obtained in discretized Brillouin zones following the method in Ref.~[\onlinecite{Fukui_2005}]. 

\section{Evaluation of the Chern number in terms of effective masses}
\label{app_effective_mass}

It is well known that with the introduction of the staggered sublattice potential $\Delta$ as mentioned before, a $C=1$ topological phase can be induced into the phase diagram of the Haldane-Hubbard model~\cite{He_2011,Vanhala_2016}. The same can happen for the Kane-Mele Hubbard model, according to Ref.~[\onlinecite{Jiang_2018}]. The physical understanding of the phase [usually known as the antiferromagnetic Chern insulator (AFCI)] is that due to the interplay between the topology and electron interactions, a spontaneous spin-rotation symmetry breaking takes place, with one of the spin components remains topologically nontrivial whereas the other does not. The conclusion has been supported by various methods, including the MF analysis, the exact diagonalization (ED) and the density matrix renormalization group (DMRG), quantum Monte Carlo, and the dynamical mean-field theory (DMFT)~\cite{He_2011,Vanhala_2016,Tupitsyn_2019,Mertz_2019,Shao_2023,He_2023}. In this section, we would like to show that in the MF level, the $C=1$ phase that emerges by introducing a sublattice-dependent on-site Coulomb repulsion, i.e., $U_{\text{A}}\neq U_{\text{B}}$, indeed shares the feature with the AFCI phase in the previous literature.

To obtain a qualitative understanding of the appearance of the AFCI phase, let us first focus on the diagonal terms in $H_{\text{I-MF}}$, ignoring the off-diagonal terms temporally. In general, these diagonal terms, which are momentum independent, can be expressed in the form as
\begin{eqnarray}
m_0 \text{I}+m_s \sigma_z +m_\theta \tau_z +m_{\theta s} \sigma_z \tau_z.
\label{eq_mass}
\end{eqnarray} 
where the Pauli matrices $\sigma$ and $\tau$ represent the spin and sublattice indices. The values of $m$'s can be uniquely determined by the four diagonal entries. If we follow the previous discussions of the noninteracting part $\hat{H}_0$ and concentrate on the mass terms of the effective Dirac Hamiltonians around the $\pm\mathbf{K}$ points, we find that the masses for the up- and down-spin components are
\begin{equation}
m_{\sigma}\left(\pm\mathbf{K}\right)=m_{\theta}+m_{\theta s} \sigma_z \pm 3\sqrt{3} t_2\sin\phi.
\label{eq_msigma}
\end{equation}
In our calculation, we have chosen the parameters $t_2$ and $\phi$ fixed to be $0.2$ and $\pi/2$, respectively. We see that if $m_{\theta s}\neq 0$, the effective masses for the up- and down-spin components ($\sigma_z=\pm 1$) can be different. It leaves room for the possibility of driving one spin component into the topologically trivial state while the other remains in the nontrivial state (the signs of the masses remaining opposite for the $\pm\mathbf{K}$ points). The possible outcomes can be summarized as follows:
\begin{eqnarray}
&&\left(\operatorname{sgn}\left(m_{\sigma}(\mathbf{K})\right), \operatorname{sgn}\left(m_{\sigma}(-\mathbf{K})\right)\right) \nonumber\\
&=&\left\{\begin{array}{lcl}(+1,+1), && C_{\sigma}=0; \\ (-1,-1), && C_{\sigma}=0; \\ (+1,-1), && C_{\sigma}=1; \\ (-1,+1), && C_{\sigma}=-1.\end{array}\right.
\label{eq_spinchern}
\end{eqnarray}
Then the total Chern number simply reads $C=C_{\uparrow}+C_{\downarrow}$ if all the off-diagonal terms in $\hat{H}_{\text{I-MF}}$~(\ref{eq_himf}), including the spin-flipping terms, are dropped.

To be more specific, we can define a MF spin-dependent staggered potential as 
\begin{equation}
\Delta^{\sigma}_{\text{MF}}=m_{\theta}+m_{\theta s} \sigma_z,
\label{eq_deltasigma}
\end{equation}
and the above expression for the masses~(\ref{eq_msigma}) can be rewritten as $m_{\sigma}=\Delta_{\text{MF}}^{\sigma}\pm 3\sqrt{3}t_2\sin\phi$. It is easy to show that in terms of the MF parameters in the MF Hamiltonian~(\ref{eq_himf}), $\Delta^{\sigma}_{\text{MF}}=\left(\varepsilon^{\text{A}}_{\sigma}-\varepsilon^{\text{B}}_{\sigma}\right)/2$. The nonzero $\Delta_{\text{MF}}$ tends to smear out the sign difference of the mass terms between the two valleys. The inequality of $\Delta_{\text{MF}}^{\uparrow}$ and $\Delta_{\text{MF}}^{\downarrow}$, if it happens, indeed indicates the breaking of the spin symmetry and associated with that, the possibility of the emergence of the AFCI phase (with $C=1$). The disparity between $\Delta_{\text{MF}}^{\uparrow}$ and $\Delta_{\text{MF}}^{\downarrow}$ is produced by $m_{\theta s}$, which in terms of the MF particle densities, reads
\begin{equation}
m_{\theta s}=\frac{1}{2}\left(U_{\text{B}}S^{\text{B}}_z-U_{\text{A}}S^{\text{A}}_z\right),
\label{eq_mthetas}
\end{equation}
where the spin polarization $S^{\alpha}_z=\frac{1}{2}\left(n^{\alpha}_{\uparrow}-n^{\alpha}_{\downarrow}\right)$ with $\alpha=\text{A}/\text{B}$. Here to be complete, we also present the expression for the homogeneous part of $\Delta_{\text{MF}}^{\sigma}$, i.e., $m_{\theta}$,
\begin{equation}
m_{\theta}=\frac{1}{4}\left(U^{\text{A}}n^{\text{A}}-U^{\text{B}}n^{\text{B}}\right)-\frac{3}{2}V\left(n^{\text{A}}-n^{\text{B}}\right).
\end{equation}

Hitherto, we only demonstrate the influence of the diagonal terms of the MF Hamiltonian on the evolution of the Dirac masses. A more quantitative analysis, even within the MF level, no doubt should include the off-diagonal terms, i.e. $\varepsilon_{\sigma\sigma^{\prime}}^{\text{A/B}}$ and $\xi_{\sigma\sigma^{\prime}}(\mathbf{k})$, which are the results of the Fock decoupling of the on-site $U$ and the NN $V$ interactions. The situation is quite complicated if we try to do so and an exact analytical treatment remains to be found. Here, however, we try to provide an approximation scheme within the frame of effective masses, and the strategy is following. 

We first diagonalize the $4\times 4$ matrix $\mathcal{H}_{\text{I-MF}}$ in Eq.~(\ref{eq_himf}) at each valley (by putting $\mathbf{k}=\pm\mathbf{K}$). With the four eigenvalues and Eq.~(\ref{eq_mass}), we can determine the effective masses. Then for each valley, the spin-dependent staggered potential can be obtained according to Eq.~(\ref{eq_deltasigma}). We denote it as $\delta_{\text{MF}}^{\sigma}$, to emphasize the fact that it is an effective staggered potential after considering the contribution from the off-diagonal terms. The sign of the effective Dirac mass can then be determined, followed by the estimation of the Chern numbers. It is worth noting that the calculation shows zero results for $m_s$ and $m_{\theta s}$ in the CDW and the $C=2$ phase; while in the $C=1$ AFCI phase and the SDW phase, both quantities are away from zero. The facts suggest the SU(2) spin symmetry breaking in these phases. 

\begin{figure}
\centering
\includegraphics[width=0.5\textwidth]{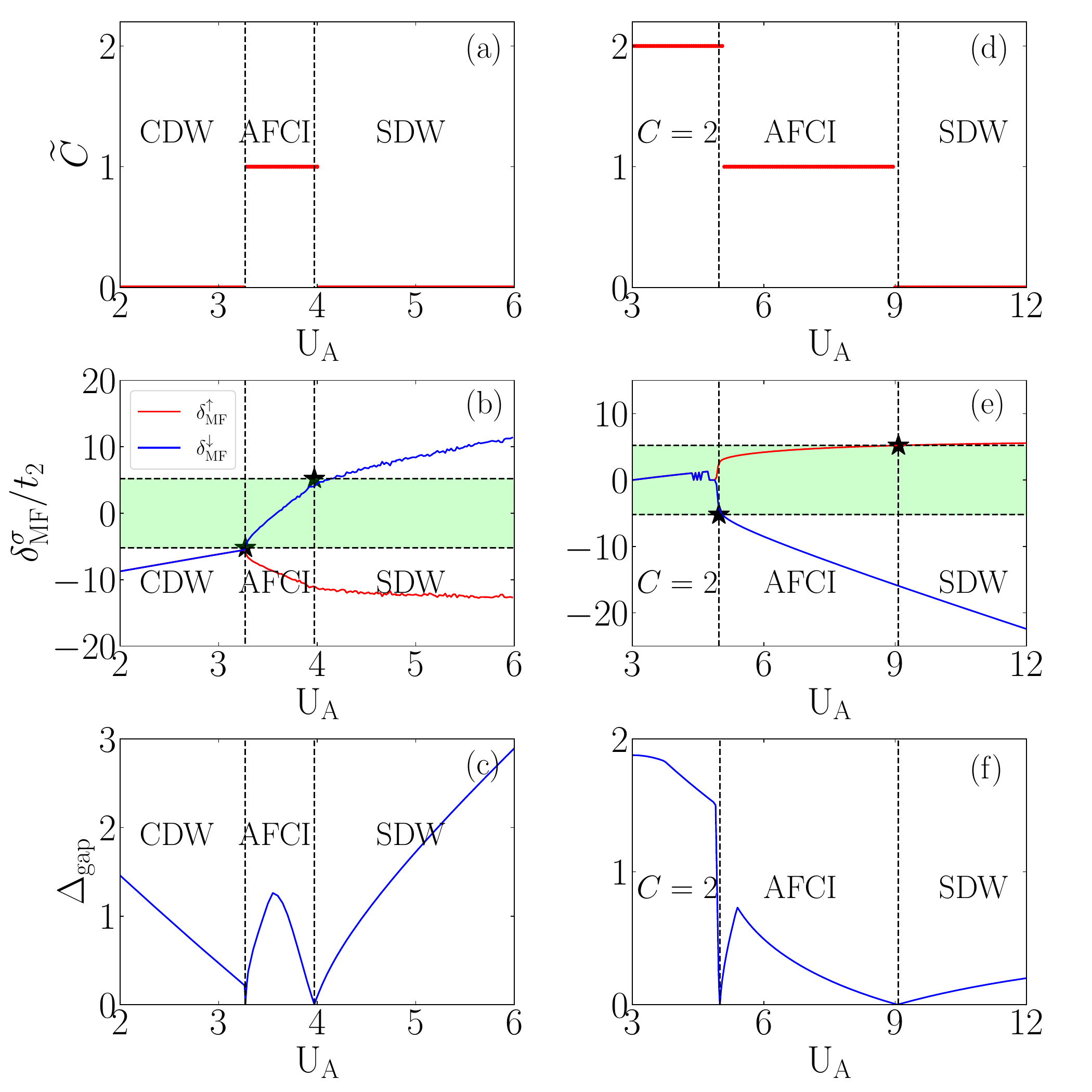}
\caption{The effective-mass analysis of the topological phase transitions in the extended Haldane-Hubbard model with sublattice-dependent repulsion. Left column for $V=1.0$, $U_{\text{B}}=6.5$, right column for $V=0.2$, $U_{\text{B}}=3.0$. The number, denoted as $\tilde{C}$, evaluated from the effective-mass method is shown in the first row. The effective spin-dependent staggered potential $\delta_{\text{MF}}^{\sigma}$ in the unit of $t_2$ as a function of $U_{\text{A}}$ is presented in the second row. The shaded region, bounded by $\pm 3\sqrt{3}$ horizontal lines, marks the topological regime. The intersections of the curves and the boundary lines are marked by stars. The third row displays the evolution of the gap between the two lower bands (occupied) and the two upper bands (unoccupied).}
\label{fig_mass}
\end{figure}
% (see text for detail)

The results of our calculation are presented in Fig.~\ref{fig_mass} with two sets of parameters: one is $V=1.0$, $U_{\text{B}}=6.5$ (left column); the other is $V=0.2$, $U_{\text{B}}=3.0$ (right column). With increase of $U_{\text{A}}$, the system experiences three consecutive phases, namely, CDW-AFCI-SDW and ($C=2$)-AFCI-SDW, respectively. In each subfigure, the phase transition points are indicated by vertical dashed lines, which are determined by the Chern number via the method in Ref.~[\onlinecite{Fukui_2005}]. They are accurate for a given band structure. From the last row in Fig.~\ref{fig_mass} [Figs.~\ref{fig_mass}(c) and \ref{fig_mass}(f)], we see that at each transition point, accompanied by the change of the Chern number, the band gap (the gap between the two upper bands and the two lower bands) always closes. The first row in Fig.~\ref{fig_mass} [Figs.~\ref{fig_mass}(a) and \ref{fig_mass}(d)] presents the Chern number (denoted as $\tilde{C}$) in terms of the sign change of the effective masses as detailed before. The second row [Figs.~\ref{fig_mass}(b) and \ref{fig_mass}(e)] shows the evolution of the effective staggered potential for each spin component (denoted as $\delta_{\text{MF}}^{\uparrow/\downarrow}$) as a function of $U_{\text{A}}$. The shaded region, bounded by $\pm 3\sqrt{3}$ horizontal lines, marks the topological regime. From the results, we see that although it is not an exact solution, the method of the effective masses works quite well, except for the region very close to the phase boundaries. 

Before concluding this section, we would like to add some remarks. In the MF analysis of the ionic Haldane-Hubbard model~\cite{He_2023}, where the effective spin-dependent staggered potentials were also calculated, it showed that the approximation with {\em only} diagonal terms of the MF Hamiltonian works quite well. However, in our case of the extended Haldane-Hubbard model with sublattice-dependent repulsion (without the ionic term), we find that the off-diagonal terms need to be considered to produce decent results. 
% Among these terms, we also notice that $\xi_{\sigma\sigma^{\prime}}$ is not essential for the final results and can be largely omitted. 

% \bibliographystyle{apsrev4-2}
% \bibliography{lt.bib}

%

\end{document}